\documentstyle[12pt,draft,epsf]{article}
\newcommand{\Tr}{\mbox{Tr}}
\begin{document}

\title {Glue Ball Masses and the Chameleon Gauge}

\author{E. Marinari$^{(a)}$, M. L. Paciello$^{(b)}$,
        G. Parisi$^{(c)}$\\
   and B. Taglienti$^{(b)}$\\[0.5em]
  {\small (a): Dipartimento di Fisica and Infn, Universit\`a di Cagliari}\\
  {\small \ \  Via Ospedale 72, 09100 Cagliari (Italy)}\\
  {\small (b): Infn Sezione di Roma,}\\
  {\small \ \  P. A. Moro 2, 00185 Roma (Italy)}\\
  {\small (c): Dipartimento di Fisica and Infn, Universit\`a di Roma
    {\em La Sapienza}}\\
  {\small \ \  P. A. Moro 2, 00185 Roma (Italy)}\\[0.3em]
   {\small \tt marinari@ca.infn.it}\\
   {\small \tt paciello@roma1.infn.it}\\
   {\small \tt parisi@roma1.infn.it}\\
   {\small \tt taglienti@roma1.infn.it}\\[0.5em]}

\date{October 5, 1995}

\maketitle
\begin{abstract}
We introduce a new numerical technique to compute mass spectra, based on
difference method and on a new gauge fixing procedure. We show that the
method is very effective by test runs on a $SU(2)$ lattice gauge theory.
\end{abstract}
\vfill
\begin{flushright}
  { \tt hep-lat/9511012 }
\end{flushright}

\newpage

Even if computations of the gluonic mass spectrum have reached
recently very high levels of precision and reliability \cite{LAT}
looking for more effective physical procedures to measure correlation
functions is a very important task. The measurement of connected
correlation functions from Monte Carlo procedures is a very time
consuming task, and the interesting signal comes from large
separation, where the functions we want to determine are exponentially
small.

In this note we will continue in exploring the possibility of using {\em
continuous} updating schemes (Langevin-like, as opposed to the discrete
step Metropolis scheme) and to exploit such continuity to define
connected correlation functions by means of differences of correlated
dynamical processes \cite{PARISI,ONE} (for analytic applications of these
kinds of methods see for example ref. \cite{PARMA}).
Here we will propose a new method which takes care about
a crucial ingredient, the gauge invariance of the theory, and that looks
far more effective than the pre-existing schemes.

The simple idea of the difference scheme to compute connected correlation
functions is the following. In the rest of the paper we will
consider the $0^{++}$
glueball mass, defined from the connected correlation function

\begin{equation}
	G(t) \equiv \langle E^{(3)}(0) E^{(3)}(t)\rangle -
	\langle E^{(3)}(0) \rangle   \langle E^{(3)}(t)  \rangle
	\simeq \exp\{-m_{0^{++}}t\}\ ,
	\label{E_GT}
\end{equation}
where with the upperscript $3$ we denote the sum of the spatial plaquettes
on the $3$-cube at a given time ($0$ and $t$ in the previous equation).
We will consider a system governed by Wilson Action at inverse
temperature $\beta=T^{-1}$, i.e.

\begin{equation}
	S_\beta \equiv \beta \sum_{plaquettes} (1-\frac12 \Tr\  U_P) \ ,
	\label{E_WIL}
\end{equation}
for $SU(2)$ gauge fields, with the usual definition of the plaquette
variables $U_P$.

A possibly effective way to measure connected correlation functions has
been suggested \cite{PARISI,ONE}, and it is based on simulating the
dynamics of two copies of the system, one with the original action
(\ref{E_WIL}) and one with a modified action. In our case we only modify
the Action involving links on the $3d$ cube at $t=0$ (we are in $4$
euclidian space-time dimensions, and we identify one of these dimensions
with the euclidian time). So, for $t\ne 0$ we use the original Wilson
Action, while at $t=0$ we modify the value of the coupling, by setting

\begin{equation}
	\beta \ \to\  \beta\  +\  \delta\beta\ ,
	\label{E_DELTABETA}
\end{equation}
with a small $\delta\beta$. In the limit of $\delta\beta\to 0$ one gets
that \cite{ONE} (we indicate with a tilde expectation values taken over
the modified action)

\begin{equation}
	\tilde{E}^{(3)}(t) - E^{(3)}(t) =
	G(t) \delta\beta + O(\delta\beta^2)\ ,
	\label{E_DIFFERENZA}
\end{equation}
where by $E^{(3)}(t)$ we denote the average of the space-like plaquettes
at time $t$. In words we consider the difference of the time $t$ energy
of the unperturbed and the perturbed system, in the limit of a small
perturbation. This difference, when measured with good statistical
precision, gives us a measurement of the connected correlation function
$G$.

The main point of the method is that if we use an appropriate simulation
technique $E$ and $\tilde{E}$ are correlated, and by exploiting that we can
eliminate the most part of the statistical error we would get in a direct
measurement of the $G(t)$.  The typical pattern of such a joint simulation
is the following.  We start from two copies of the same configuration.
Their {\em energy distance} is zero, since we are out of equilibrium.  Now
we start the two simulations.  The energy distance at a given time $t$ will
hopefully stabilize after a transient period ($\tau_0$, we indicate with
$\tau$ the dynamical fifth time, as opposed to euclidian time $t$) at the
correct distance.  Then for large dynamical time ($\tau>\tau_1$) the two
trajectories in phase space will separate, the statistical noise will
dominate the signal and the measurement will be of little use.  The
measurement window  goes from $\tau_0$ to $\tau_1$, and a good dynamical
procedure maximizes its extent.

A discrete dynamics like the Metropolis algorithm does not do the job.
Indeed in this case the advantage of the method, which to be well
performing exploits the fact that the two trajectories are close in phase
space, is lost due to the intrinsic discreteness of the updating procedure.
On the contrary a Langevin dynamics can be the basis of our scheme
\cite{PARISI,ONE}. For the two copies of the system we write

\begin{eqnarray}
    \dot{U} & = & -\frac{\delta S}{\delta U} + \eta\ , \\
	\dot{V} & = & -\frac{\delta S}{\delta V} + \eta\ , \nonumber
\end{eqnarray}
where we have denoted by $U$ the unperturbed fields and by $V$ the
perturbed ones, and $\eta$ is the same noise for the two systems.

For a gauge model the gauge degrees of freedom create an additional
complication. The gauge part of the degree of freedom random walks in
phase space, and such a random walk tends to separate the two
trajectories. An usual gauge fixing (for example putting to one all
time-like gauge variables) does slow down the dynamics making the method
inpractical \cite{ONE}. In ref. \cite{ONE} we have seen that sometimes the
phenomenon can be dramatic enough to make any measurement impossible
(even for short euclidean time separations).

In ref.  \cite{ONE} we proposed to solve the problem by using some kind of
magnetic field, which would insure a smooth, partial fixing of the gauge.
Wilson action would be modified by a non-gauge invariant term, selected in
such away to have a small effect one gauge-invariant quantities.  Even if
this method was improving the situation, it did not turned out to be very
superior to usual methods.  Trajectories in phase space did diverge quite
soon, and a slow drift on the system internal energy (that is also modified
of a small amount due to the magnetic like term) was difficult to control.
Also a non-gauge invariance formulation is definitely not so appealing, and
could present a large number of unwanted features.

In this note we propose a new method, and we show that it is indeed very
effective. The basic idea is very simple. We fix the gauge where the two
configurations are as similar as possible. We call such a gauge the {\em
chameleon gauge}. This method does indeed keep the gauge part of the two
systems as close as possible (reducing the rate of divergence of the two
trajectories in phase space) but does not introduce any sizeable slowing
down in the dynamics of observable quantities like the energy.

After each full lattice sweep of Langevin update of the
link variables $U$ and $V$ we gauge fix the $U$ configuration (obviously
it does not change to gauge fix instead the $V$ configuration). We
maximize the quantity

\begin{equation}
	\sum_{n,\mu} U_\mu(n) V^\dagger_\mu(n) \ ,
	\label{E_MAXIMIZE}
\end{equation}
where $N$ runs over the lattice sites and $\mu$ over the $4$ directions.
We find the gauge transformation $\{  g(n)  \}$, with

\begin{equation}
	U_\mu(n) \to g(n) U_\mu(n) g^\dagger(n+\hat{\mu}) \ .
	\label{E_GAUGETRA}
\end{equation}
We determine $\{  g(n)  \}$ such that

\begin{equation}
		\sum_{n,\mu} g(n) U_\mu(n) g^\dagger(n+\hat{\mu}) V^\dagger_\mu(n)
		= \mbox{local maximum}\ .
	\label{E_LOCALMAX}
\end{equation}
The quantity

\begin{equation}
	{\cal F} \equiv \frac{1}{4V}\sum_{\mbox{links}}
	(1- U_\mu(n) V^\dagger_\mu(n))\ ,
	\label{E_MONITOR}
\end{equation}
gives a measure of how good a gauge fix we have been able to reach.
${\cal F}=0$ when the two configurations are identical. A high value of
${\cal F}$ signals that a good gauge fix has not been achieved (implying
for example that the two gauge field configuration differ because of a
physical, non gauge freedom related reason).

\begin{figure}
        \epsfxsize=400pt\epsffile{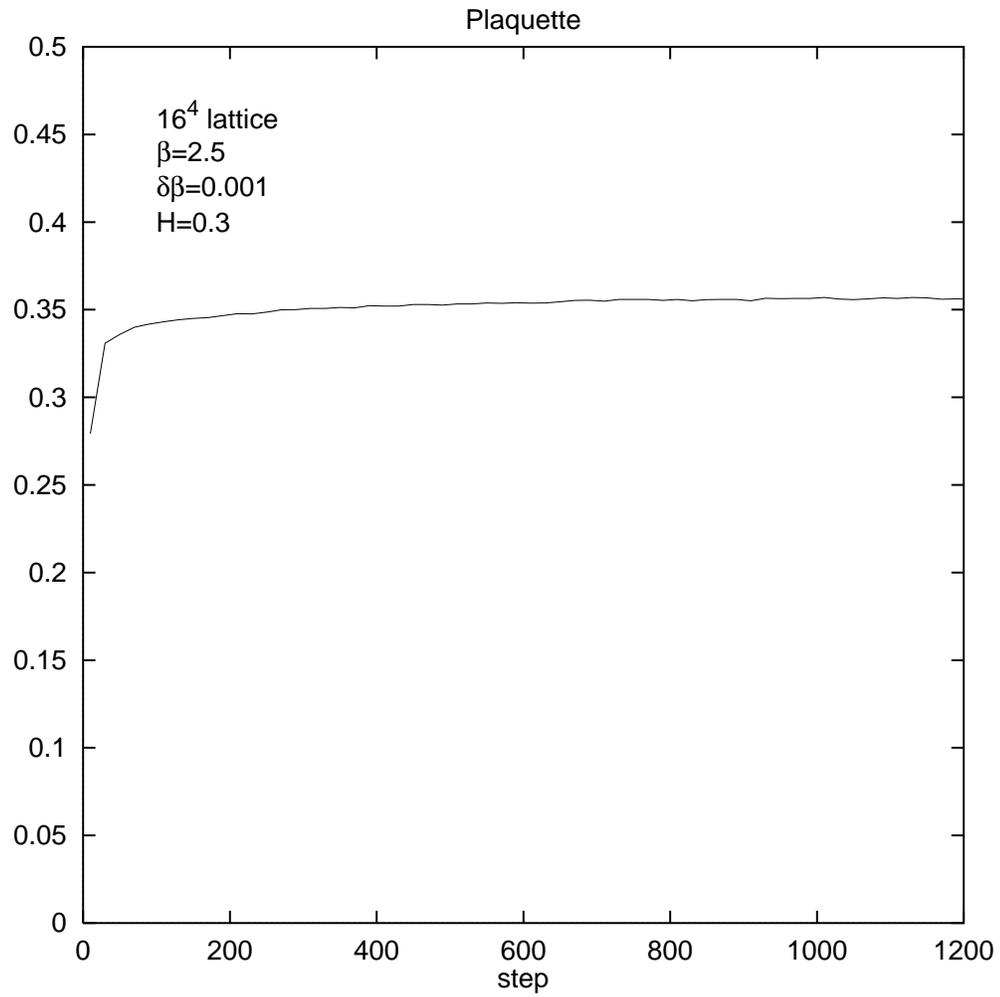}
	\caption[1]{
        The plaquette operator expectation value computed by using
        the magnetic field method.}
	\protect\label{F_FIGE1}
\end{figure}

\begin{figure}
        \epsfxsize=400pt\epsffile{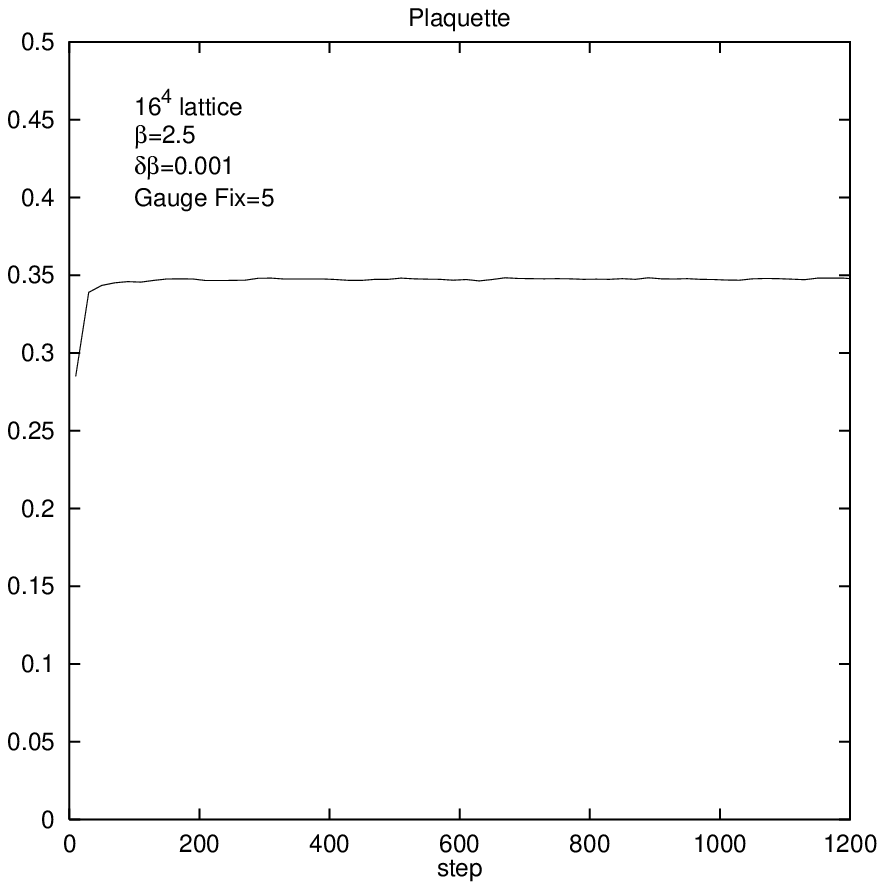}
	\caption[1]{
        The plaquette operator expectation value
        computed by using the chameleon gauge fixing method.}
	\protect\label{F_FIGE2}
\end{figure}

Our numerical simulations show that the method is very well
performing, and has very pleasant features. The (correct) equilibrium
value for the plaquette energy is reached in very short time. That
means that all the problems connected to standard gauge fixing (like
axial gauge) or to our magnetic field partial gauge fixing have been
solved by the chameleon gauge.  We show in fig. (\ref{F_FIGE1}) the
plaquette value computed with the magnetic field method, where a slow
drift is evident. In fig. (\ref{F_FIGE2}) we show the new data, where
after less than $100$ Monte Carlo sweeps the measured value is stable.
Correlation functions measured by difference technique on the
chameleon gauged configurations are far less noisy than with our old
method. In a typical situation we gain a factor of order $5$ over the
time window we can rely on. It is also quite interesting to note that
the breakdown of the correlation functions (that become noisy after a
given time $\tau_1$) is always signaled by a sudden growth of the
quantity ${\cal F}$, i.e. by a collapse of the quality of the gauge
fixing we are able to reach.

Let us give a few details about our runs and our numerical results. We
simulate an $SU(2)$ system on a $16^4$ lattice with periodic boundary
conditions. We keep one copy of the system at the original $\beta$ value,
while we simulate four parallel systems, where different time slices are
set at the inverse square coupling value $\beta+\delta\beta$. The program
size is $44$ Mbyte, and on a IBM RISC WS 550E we run $500$ steps in close
to one day. We have selected $\beta=2.5$, that is close enough to the
scaling region, and where our lattice size is large enough to describe
infinite volume behavior. Michael and Teper estimate in ref.
\cite{MICTEP} that $ma \simeq .660$. The result we obtain by using our
method is fully compatible with that, and from a global fit to our data
we obtain $ma \simeq .67$, with an error, including both statistic
uncertainty and systematic effects from finite euclidean time distance
that we estimate to be smaller than $5\%$.

\begin{figure}
        \epsfxsize=400pt\epsffile{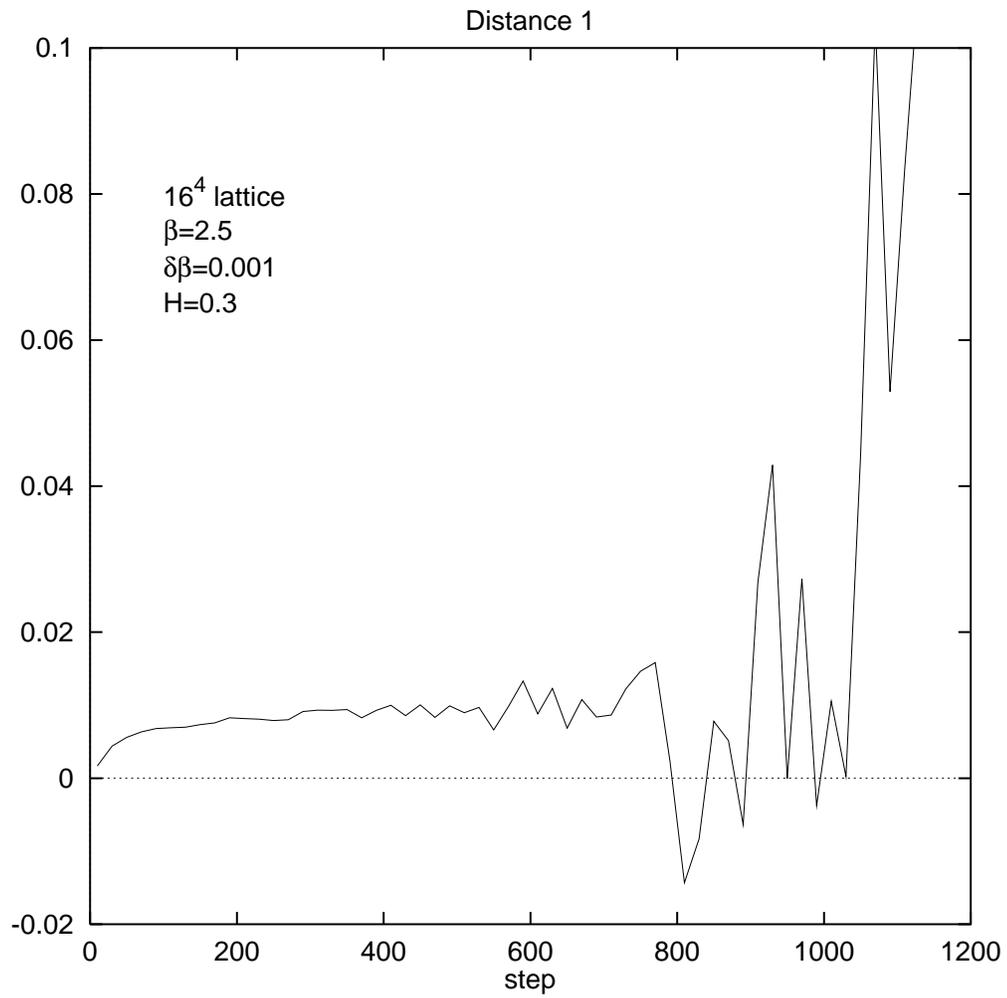}
	\caption[1]{
        The distance $1$ correlation computed by using the magnetic
	field method.}
	\protect\label{F_FIG1}
\end{figure}%

\begin{figure}
        \epsfxsize=400pt\epsffile{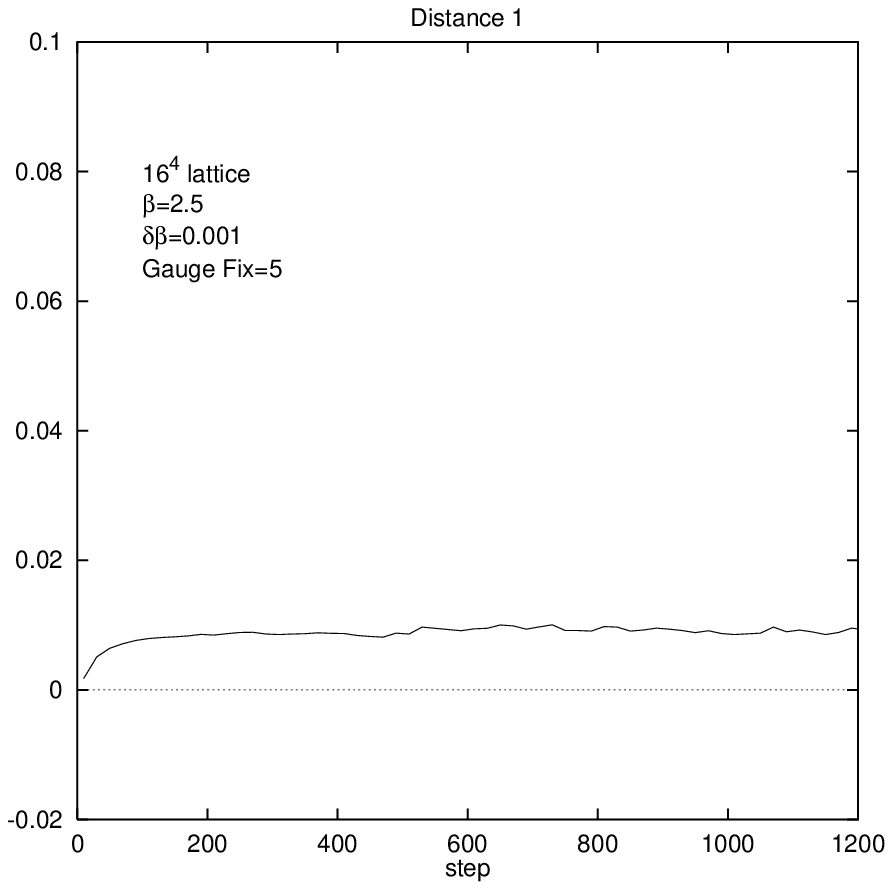}
	\caption[1]{
        The distance $1$ correlation computed by using the chameleon
	gauge fixing method.}
	\protect\label{F_FIG2}
\end{figure}%

In fig. (\ref{F_FIG1}) we show the signal we were getting  at separation
$1$ by using our old method (based on the use of the magnetic gauge
fixing term), while in fig. (\ref{F_FIG2}) we show the result we obtain
with our new method. The difference is quite impressive. While signal was
becoming noisy at $\tau \simeq 400$ and was lost at $\tau \simeq 800$ in
the old approach, in the chameleon gauge fixed approach we get a stable
signal up to time $\tau > 1200$.

We believe that there is still much to understand in the physics of the
so-called {\em difference methods}, both in gauge and in simple spin
systems, but we hope that the method we are proposing here is a nice step
in the direction of less expensive measurements of mass spectra.

\end{document}